\def\hh{H$_2$}
\def\hhhp{H$_3^+$}
\def\ohp{OH$^+$}
\def\hhop{H$_2$O$^+$}
\def\hhhop{H$_3$O$^+$}
\def\hho{H$_2$O}
\def\hnop{H$_n$O$^+$}
\def\meth{CH$_3$OH}
\def\ammo{NH$_3$}
\def\pow#1#2{#1$\times$10$^{#2}$}
\def\scm{cm$^{-2}$}

\documentclass[
    ,final            
  ]
  {aipproc}

\layoutstyle{8x11single}

\begin{document}

\title{The chemistry of interstellar \hnop: Beyond the Galaxy}


\author{Floris van der Tak}{  address={SRON Netherlands Institute for Space Research, Landleven 12, 9747 AD Groningen, The Netherlands}
}


\begin{abstract}
 The astrochemistry of the \hnop\ (n=1..3) ions is important as the main gas-phase formation route for water, and as tracer of the interstellar ionization rate by cosmic rays and other processes. While interstellar \hhhop\ has been known since the early 1990's, interstellar \ohp\ and \hhop\ have only recently been detected using the Herschel space observatory and also from the ground. This paper reviews detections of \hnop\ toward external galaxies and compares with ground-based work. The similarities and differences of the H$_n$O$^+$ chemistry within the Galaxy and beyond are discussed.
\end{abstract}

\maketitle

\section{Introduction}

The field of astrochemistry is a rich and rapidly growing field: as of Summer 2010, over 150 molecular species are known to occur in interstellar and circumstellar environments. Most of these species are simple di- and triatomic radicals and ions, but in regions shielded from stellar ultraviolet radiation, polyatomic molecules are able to survive as well. Molecular complexity in space reaches its peak in the so-called hot cores, where the products of surface chemistry on small solid particles (dust grains) evaporate into the gas phase, leading to organic species such as simple sugars and alcohols.
While the astronomers' definition of a complex molecule (more than 4 atoms) is quite far from the chemists' one, the observations show that considerable molecular richness is built up in space despite the rather unfavourable conditions. See \citet{herbst:ewine} for a review of the subject.

Oxygen is the third most abundant element in the Universe, after hydrogen and helium, and the two most abundant oxygen-bearing molecules are CO and \hho. The formation and destruction of CO are well understood: in dense molecular clouds, all gas-phase carbon is locked up in CO, while at low (column) densities, photodissociation limits the CO abundance. The main chemical activity of CO in dense clouds is its freeze-out onto grain surfaces at low temperatures and high densities, and its subsequent evaporation into the gas phase when the grains are warmed up. In contrast, the formation and destruction of \hho\ is an interplay between gas-phase and grain surface reactions and photodissociation. This dependence makes \hho\ an excellent tracer of astrophysical conditions, in particular instances when energy is injected into interstellar gas clouds.

The gas-phase formation of \hho\ proceeds via two channels. In the neutral-neutral route, O and OH react with \hh\ to form OH + H and \hho\ + H, respectively, but both reactions have activation barriers so that this mechanism is only effective at gas temperatures above $\approx$250\,K. At lower temperatures, the ion-molecule route dominates, starting with the reaction of O with \hhhp\ or by charge exchange between O and H$^+$. Reactions of O$^+$, \ohp\ and \hhop\ with \hh\ produce \hhhop, which recombines with an electron to form \hho\ + H, as well as OH + \hh. While observational evidence for this route existed from ground-based observations of \hhhop\ \citep{phillips:h3o+}, direct evidence for the intermediate products \ohp\ and \hhop\ was lacking until the launch of ESA's \textit{Herschel} space observatory.

\section{Extragalactic \hnop\ with Herschel}

The nucleus of the galaxy M82 is well known for its high rate of star formation. To measure the abundance of \hho\ in M82, \citet{weiss:m82} have obtained spectra of the \hho\ $1_{10}-1_{01}$, $1_{11}-0_{00}$ and $2_{02}-1_{11}$ transitions with the high-resolution HIFI spectrometer onboard Herschel. Unexpectedly, the spectrum near 1113\,GHz shows absorption in the \hhop\ $1_{11}-0_{00}$ line at 1115\,GHz, which is stronger than the neighbouring \hho\ line. The authors infer column densities of \pow{4}{14}\,\scm\ for \hho\ and \pow{3}{14}\,\scm\ for \hhop. This high relative abundance of \hhop\ is surprising as \hhop\ reacts quickly with \hh\ to form \hhhop. The observations thus indicate that a significant fraction of hydrogen is in atomic form, unless another process is reducing the \hho\ abundance.

Even more extreme conditions are found in the galaxy Mrk 231, where a supermassive black hole is probably responsible for the high infrared luminosity. An FTS spectrum of the Mrk 231 nucleus, taken with SPIRE on Herschel, shows emission in high-$J$ lines of CO and \hho, in the \hhop\ 1115\,GHz line, and in the \ohp\ $1_0-0_1$, $1_2-0_1$ and $1_1-0_1$ lines near 909, 972, and 1033 GHz \citep{vdwerf:mrk231}. While the ISO-LWS spectra of Mrk 231 and Arp 220 show hints of \ohp\ lines \citep{gonzalez:mrk231}, the \hhop\ detection is the first outside the Solar System, where Herzberg identified \hhop\ in cometary spectra \citep{herzberg:h2o+}. The high brightness of the CO lines in Mrk 231 arising from levels up to 460\,K above ground, and the significant lower limits of $\sim$\pow{2}{-10} for the \ohp\ and \hhop\ abundances indicates that X-ray emission from the black hole vicinity is driving the heating and the chemistry in the nucleus of Mrk 231.

\section{Extragalactic \hnop\ from the ground}

\begin{figure}
  \includegraphics[height=.5\textheight]{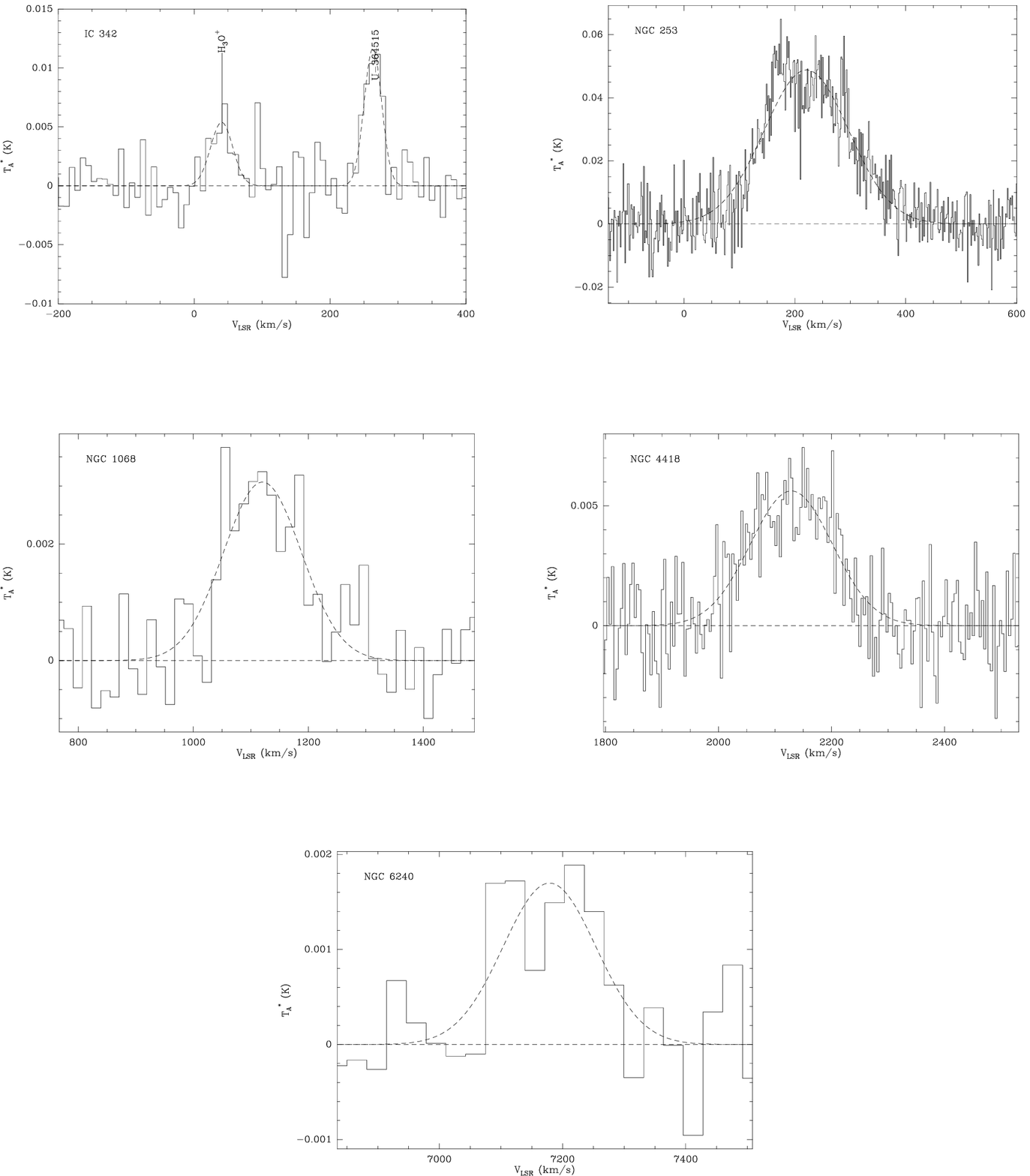}
  \caption{Recent detections of \hhhop\ with the JCMT \citep{aalto}.}
\end{figure}

Whereas \hhop\ cannot be observed from the ground, detection of \ohp\ has been achieved with the APEX telescope \citep{wyrowski:oh+}, although not yet toward extragalactic objects. In contrast, the first detection of extragalactic \hhhop\ by \citet{vdtak:h3o+} is now being followed up in several ways (Figure 1). 
While \citet{aalto} concentrate on the 364\,GHz line in a larger source sample,
the line survey by \citet{requena} covers two lines of \hhhop\ to constrain the excitation of \hhhop, which reduces the uncertainty of the column density estimates. The blend of the 307\,GHz line with a \meth\ line can be resolved since their broad-band spectra include $>$10 other \meth\ lines.


The estimate of $N$(\hho) toward the nucleus of the starburst galaxy M82 from \citet{weiss:m82} exceeds the estimate of $N$(\hhhop) from \citet{vdtak:h3o+} by a factor of only $\approx$3.3, which is
incompatible with models of pure gas-phase chemistry, even under strong irradiation by ultraviolet photons or X-rays. Models of X-ray irradiated gas such as expected in the vicinity of a supermassive black hole predict \hho/\hhhop\ ratios of $\sim$100, while models of UV-irradiated gas around young massive stars predict even higher ratios: \hho/\hhhop$\sim$1000 for a young starburst, dropping to $\sim$300 after $\sim$10\,Myr, when supernova explosions start enhancing the local ionization rate by energetic particles (`cosmic rays'). 

Additional ionizing radiation to increase the production rate of \hhop\ seems unavailable, and increasing the atomic fraction of the gas is problematic since \ohp\ and \hhop\ require \hh\ for their formation too.
Therefore the solution to the low observed \hho/\hhhop\ ratio in M82 may be an enhanced destruction rate of \hho. The photodissociation rate of \hho\ is $\approx$70$\times$ higher than that of \ohp\ \citep{ewine:photo,roberge:photo}, while the rate for \hhop\ is negligible \citep{ewine:fd}.  
The rate for \hhhop\ may be high like \hho\ and \ammo, to which it is iso-electronic, or low like that of \ohp\ and \hhop.
In the latter case, a scenario is possible where \hho\ molecules from icy grain mantles are ejected into the gas phase by shocks and subsequently photodissociated into O and OH. The dissociation products then participate in `standard' ion-molecule chemistry (\S\,1) which produces \hhhop. \citet{weiss:m82} use similar arguments to explain the observed \hhop/\hho\ ratio in M82. 
This production mechanism of \hhop\ is similar to that in comets, where \hhop\ is made by photoionization as well as electron impact ionization of evaporating \hho\ \citep{bhardwaj:comet}. 
In the case of M82, shocks are more likely to cause the grain mantle evaporation than stellar radiation which only heats small volumes of gas. Shock waves pervade the nucleus of M82 as a result of cloud-cloud collisions due to the shape of the stellar bar potential \citep{greve:m82}. However, the \hho/\hhop\ ratio in Comet Halley is $\sim$10$^5$ \citep{bhardwaj:comet} and it is not clear if the different radiation field can explain the difference with M82. Testing the photoevaporation-ionization hypothesis for \hhhop\ requires measurement or calculation of its photodissociation cross-section.

\section{Comparison with the Galactic case}

The \ohp\ and \hhop\ ions have been detected on a number of Galactic lines of sight: the 971\,GHz line of \ohp\ and the 607 and 1115\,GHz lines of \hhop\ with the \textit{Herschel} telescope \citep{gerin:oh+,ossenkopf:h2o+} and the 909\,GHz line of \ohp\ from the ground with the APEX telescope \citep{wyrowski:oh+}. In all Galactic cases, the lines appear in absorption toward a background source of dust continuum emission, even if the neighbouring \hho\ 1113\,GHz line appears in emission \citep{wyrowski:h2o+}.

The nucleus of M82 follows the behaviour of many Galactic lines of sight in that the \hho\ and \hhop\ lines at 1113 and 1115 GHz both appear in absorption, implying a low excitation of the molecules.
This condition can be quantified as $T_{\rm ex} < T_{\rm bg}$, where the excitation temperature $T_{\rm ex}$ is defined through the Boltzmann equation and the radiation temperature of the background $T_{\rm bg}$ is defined through the Rayleigh-Jeans law. 
A low excitation temperature is expected for the \ohp\ and \hhop\ lines, because of the high line frequency ($\nu \sim 1$\,THz) and the significant molecular dipole moments ($\mu$=2.3 and 2.4 D, respectively). The spontaneous decay rate $A_{ul}$ scales as $\mu^2 \nu^3$, so that any excitation mechanism would have to be very fast ($\sim$10$^{-2}$\,s$^{-1}$) to be able to compete with radiative decay. 
The situation is different for the extragalactic \hhhop\ detections in emission: the lower line frequency ($\approx$400\,GHz) reduces both the radiative decay rate and the intensity of the dust continuum background, which scales as $\nu^\alpha$ with $\alpha$ between 2 (optically thick case) and 4 (optically thin case). In addition, the excitation of \hhhop\ may be enhanced by absorption of far-infrared photons, an effect known as 'pumping'.

Perhaps the most surprising aspect of the Mrk 231 observations is that the \ohp\ and \hhop\ lines appear in emission, which implies a very high excitation rate as discussed above. The same result holds for the HF $J$=1$\to$0 line, which appears in emission toward Mrk 231, unlike all Galactic lines of sight \citep{neufeld:hf}. These molecules all trace interstellar gas where a significant fraction of the hydrogen is in atomic rather than molecular form, so that three excitation mechanisms are possible: collisions with H, collisions with \hh, and pumping by infrared photons.

Estimating the relative importance of collisional and radiative excitation requires collisional rate coefficients, which do not presently exist for \ohp\ and \hhop. For HF such data do exist \citep{reese:hf-he} and a calculation of its excitation may be useful as a guide. Unlike HF, the excitation of \ohp\ and \hhop\ needs a non-equilibrium treatment, as their timescales for formation, excitation and destruction are similar.
In summary, understanding the chemistry of extragalactic \hnop\ requires several sets of basic molecular data: photodissociation rates for \hhhop, and inelastic de-excitation rate coefficients for \ohp\ and \hhop\ with H, \hh\ and e as collision partners.
Searches for \ohp\ and \hhop\ emission within the Galaxy will also be useful: 
the existing pointed observations are biased toward strong continuum sources but future larger-scale mapping observations may reveal Galactic sources of \ohp\ and \hhop\ emission.

\begin{theacknowledgments}
The author thanks Ewine van Dishoeck and Rowin Meijerink (Leiden), Susanne Aalto and John Black (Onsala), Friedrich Wyrowski (Bonn) and Geoff Duxbury (Glasgow) for useful discussions.
\end{theacknowledgments}

\bibliographystyle{aipproc}   
\bibliography{ions}

\end{document}